\documentclass[%
 reprint,
 amsmath,amssymb,
 aps,
]{revtex4-1}

\usepackage{graphicx}
\usepackage{dcolumn}
\usepackage{bm}

\newcommand{\av}[1]{\langle #1\rangle}
\newcommand{\st}{s}
\newcommand{\yw}{\alpha}

\newcommand{\wre}{w}
\newcommand{\trev}{{\cal T}}
\newcommand{\GG}{\Gamma}
\newcommand{\sti}{\GG_0}
\newcommand{\stia}{\GG_{0\yw}}
\newcommand{\stib}{\GG_{0\beta}}
\newcommand{\stf}{\GG}
\newcommand{\stfa}{\GG_{\yw}}
\newcommand{\stfb}{\GG_{\beta}}
\newcommand{\Fia}{F_{0\yw}}
\newcommand{\Ffa}{F_{\yw}}
\newcommand{\GF}{{\cal F}_{\yw}}
\newcommand{\GFav}{{\cal F}}
\newcommand{\W}{W}

\newcommand{\avE}{\langle \Delta E\rangle}
\newcommand{\avD}{\langle D\rangle}

\newcommand{\avdE}{\langle \W\rangle}
\newcommand{\e}{e}

\begin{document}

\preprint{APS/123-QED}

\title{Work as a Memory Record}

\author{Miroslav Hole\v{c}ek}
 \email{holecek@rek.zcu.cz}
\affiliation{%
 New Technologies Research Center, University of West Bohemia, Plze\v{n} 301 00,  Czech Republic
}%

\date{\today}

\begin{abstract}
The possibility of a controlled manipulation with molecules at the nanoscale allows us  to gain net work from thermal energy, although this seems to be in contradiction to the Second Law of thermodynamics. Any manipulation, however,  causes some memory records somewhere in the system's surroundings. To complete the thermodynamic cycle, these records must be reset, which   costs energy that cancels the previous gain.    
The question is, what happens when this memory (information) is recorded only in the work reservoir? Then it cannot be reset because the record means  nothing but the work gain itself (e.g., the result position of a weight in the gravity field).  Is this a violation of the Second Law?  To answer the question, 
we study in this theoretical work an exchange of energy between a physical (possibly microscopic) system that is thermalized at the beginning and another (possibly microscopic) system -- the work reservoir -- during a deterministic process in an autonomous arrangement, including also an auxiliary device controlling the process.   
This arrangement is suitable for deriving some equalities which express the Second Law in a form incorporating explicitly relevant memory records (and related information).  We use these equalities in studying  a hypothetical process including many cycles in which the only non-reset memory record is that in the work reservoir during each cycle. The results show that either the work gain is canceled in following cycles (and the work reservoir fluctuates and cannot accumulate energy), or there exists an information flow from the system (an information engine), or the system cannot work in an expected way for a purely dynamic reason (this reveals a deeper connection of the studied questions with the concept of adiabatic accessibility).                     
\end{abstract}

\maketitle


\section{Introduction}

A systematic extraction of usable energy (work) from a single heat bath in thermal equilibrium is a typical problem that has strict thermodynamic limitations given by the Second Law. In Planck's formulation, for example, \emph{it is impossible to construct an engine which will work in a complete cycle, and produce no effect except the raising of a weight and cooling of a heat reservoir} \cite{Planck1917}. The  more than 140-year-old puzzling question is, however, how to explain these limitations from a microscopic point of view \cite{Maxwell1871}. A Maxwell's demon  represents in this story an arbitrary physical structure (e.g. a mechanical device, a complex biological macromolecule, or even a living creature) that may control and manipulate individual molecules to accumulate their heat energy to a work reservoir \cite{LefRex2003}. 

A detailed analysis of activities of any functioning Maxwell's demon shows, however, that it must leave some physical changes -- records -- in the system's surroundings, e.g. in the demon itself. When the demon gains information about individual molecules (measurement) or uses this information to manipulate  them (feedback), it leads to an interaction with the system and the demon's physical state becomes correlated with the system \cite{SagUed2012}. When describing these correlations with the concept of mutual information \cite{CovTho1991}, we can describe the Second Law in a generalized form usable for nano- or microsystems when correlations between two subsystems cannot be neglected \cite{SagUed2008,SagUed2009,SagUed2010,SagUed2012b,HorVai2010,Sag2011} . This generalized Second Law fully explains that the work extracted from a single heat bath by a Maxwell demon is again consumed  in the process of resetting the records to perform a true thermodynamic cycle (producing no other effect, as written in the Planck formulation) in accord with  the Landauer principle \cite{LefRex2003,ParHorSag}. The Maxwell demon thus cannot violate the Second Law \cite{MarNorVed}. 

Results of thermodynamics involving the concept of  information explicitly -- so-called information thermodynamics  --  have been demonstrated and verified in various experiments on small fluctuating systems \cite{Toy2010,Kos2014,Kos2015, Mih2016,Cot2017} (see also a review \cite{Ciliberto2017} emphasizing the role of stochastic description of thermodynamics  \cite{Seifert2012,Sekimoto2010}). There is no doubt that the Second Law must involve an information interconnection of individual systems to be usable in the description of small systems and give a true analysis of any  arrangement of a Maxwell's demon. The question is, however, if it gives the complete solution of Maxwell's paradox. Specifically, we may imagine a situation when all records in the system surroundings are reset but there is a correlation of the system with the work reservoir. In other words, when states of the work reservoir serve as memory records. These records, however, cannot be reset -- they belong to the thermodynamic cycle since the performed work is an essential purpose of thermodynamic cycles.

\begin{figure}
\includegraphics[width=0.4\textwidth]{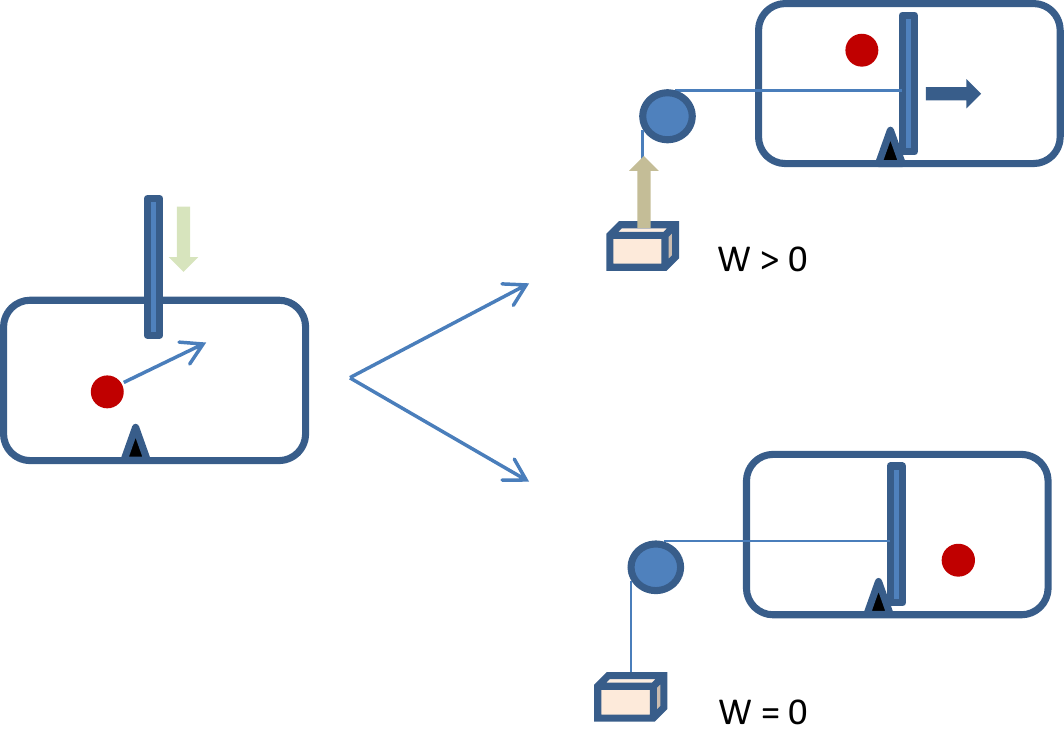}
\caption{A half-working Szilard engine that works without any measurement or feedback. It seems that in a longer time period it   may accumulate positive work. The position of the weight is a memory that records which process has passed.}
\label{fig:1}
\end{figure}

In fact, this question is very old and has been a source of much criticism \cite{JaBa1972,EaNo1999}. Imagine for instance a modification of the Szilard engine so that the piston may move only in one -- say right --  direction (see Fig. \ref{fig:1}). If the molecule is in the left part after the partition is located, it performs the work (raises a weight) without any measurement or feedback. If it is in the  right part, no work is performed since the partition does not move. So with the probability of one-half the work is not performed. Nevertheless, in a longer time period, the average positive work seems to be extracted from the heat bath. This is not in a contradiction to information thermodynamics, since the position of the weight after an individual cycle is a memory record (see also the discussion in \cite{LefRex2003}, pp. 23-25). 

In this contribution, we analyse this question on a complete autonomous supersystem involving the work reservoir as a dynamic part connected with the studied system and an auxilliary device (performing, e.g., a measurement and manipulation with the system). The supersystem is defined  in Section \ref{sec:2}, where the important equality between the performed work and the memory-dependent free energy \cite{Sag2011} is derived. In Section \ref{sec:3}, we define a complete thermodynamic process with exchange of energy with the work reservoir and a (possibly nonequilibrium) heat bath and derive some useful equalities that present, for instance, a very general form of the generalized Second Law for our arrangement. The consequences concerning the validity of the Second Law of thermodynamics are studied in Section \ref{sec:4} . In Section \ref{sec:5}, a simple model of a physical system resembling a cross over an energetic barrier is studied to show a deeper connection of the Second Law with the system's dynamic.

\section{Exchange of energy with a work reservoir \label{sec:2}}

In the thermodynamics of mesoscopic or 
microscopic  non-autonomous systems, the definition of work is connected with a change of externally controlled parameters  that vary via  a prescribed protocol \cite{Jar1997,Crooks1998,Sekimoto2010} or  as a result of the dynamics of some external macroscopic bodies whose inner degrees of freedom are ignored \cite{Peliti2008,DefJar2013}. Though these definitions of work correspond with some experimental setups (e.g \cite{Toy2010}), they cannot describe  a concrete storage (release) of energy in (from) a work reservoir as studied in some experiments, e.g. \cite{Cot2017}, and cannot be used in autonomous systems, e.g. \cite{ManJar2012,BaSei2013,Kos2015} (see also the discussion in \cite{MaTa2007}). Notice that Planck's formulation of the Second Law (or its more modern version \cite{LieYng1999}) also cannot be described within this scheme.    

We  define the work $W$ as a quantity measured via changes of a concrete (possibly microscopic) physical system called the work reservoir (WR) \cite{Shi2015,Maletal2015}. The states of WR are denoted by a variable $\wre$. The assumption concerning WR is simple: there is a function $\epsilon$ on the state space of WR so that the work accumulated in WR ($W >0$) or consumed from WR ($W<0$) during \emph{any} process in which its initial and  final states are $\wre_0$ and  $\wre$, respectively, is defined as
\begin{equation}\label{ch}
W \equiv \epsilon (\wre )-\epsilon (\wre_0).
\end{equation}
The work reservoir is thus a device whose states measure the energy submitted  into it as well as the energy transfer from WR into some other system (a typical role of a battery).

We assume also the occurrence of another system A called the \emph{auxiliary device} somewhere in the surroundings of WR. The rest of the surroundings of WR is denoted as X so that X+A+WR forms an autonomous supersystem. Both A and WR may be microscopic, mesoscopic or macroscopic systems. The auxiliary  device may not have a special role; sometime its interaction with X may be interpreted as a "measurement" on X (detection of its states) and/or a "feedback control" of X (changing its states in a demanded way). 

The crucial characterization of A and WR differentiating them from X, is that their states may be externally determined and adjusted to a given value. During a process in which the supersystem X+A+WR is isolated, an unknown state of X evolves into another state. Since the initial states of A and WR are known (they might be adjusted to  given values), their final states carry important information about the changes on X (e.g. about its change of energy).  Therefor we call  $(a,\wre )\equiv  \yw$ the \emph{memory records} of studied processes, where $a$ and  $\wre$ denote the final  state of A and WR, respectively. 
Since the memory records include the final states of WR too,  a given memory record $\yw$  defines uniquely the work performed during the process, $W\equiv W_\yw$.

Let us study various processes in which the supersystem is isolated and evolves  via its inner dynamics within a time interval with a fixed duration $\tau$. The process is deterministic, which means that the final state of X, $\st$, and its memory record are determined by its initial state $\st_0$ and the adjusted initial states of A and WR,  $(a_0,\wre_0 )\equiv  \yw_0$, i.e.
\begin{equation}\label{detevo}
\st=\st (\st_0, \yw_0),\,\,   \yw =\yw(\st_0, \yw_0).
 \end{equation}
Suppose that the initial states of A and WR at the beginning of any of the studied process have firmly given values. It implies that the final state of X depends on its initial state only, $\st (\st_0)$, if omitting the fixed initial value of the memory record $\yw_0$.

 It is useful to write the energy of the supersystem X+A+WR as consisting of two contributions: the energy of the work reservoir, $\epsilon (\wre )$, and the energy of the systems X and A including also their mutual  interaction energy and possible interaction energy of X+A with WR (if it cannot be neglected).    
Denoting $E_0$ and $E$ as the values of this rest energy at the beginning and the end of a studied process, respectively, we get the energy balance,
\begin{equation}\label{bilance}
E_0(\st_0,a_0,\wre_0)+\epsilon (\wre_0 )=
E(\st ,a,\wre )+\epsilon (\wre ). 
\end{equation}
Using the previous definitions and the energy balance, we get the equality expressing the work supplied into the work reservoir during the process, \begin{equation}\label{work}
W_\yw = E_0 (\st_0,\yw_0) - E(\st ,\yw ).
\end{equation}

\begin{figure}
\includegraphics[width=0.4\textwidth]{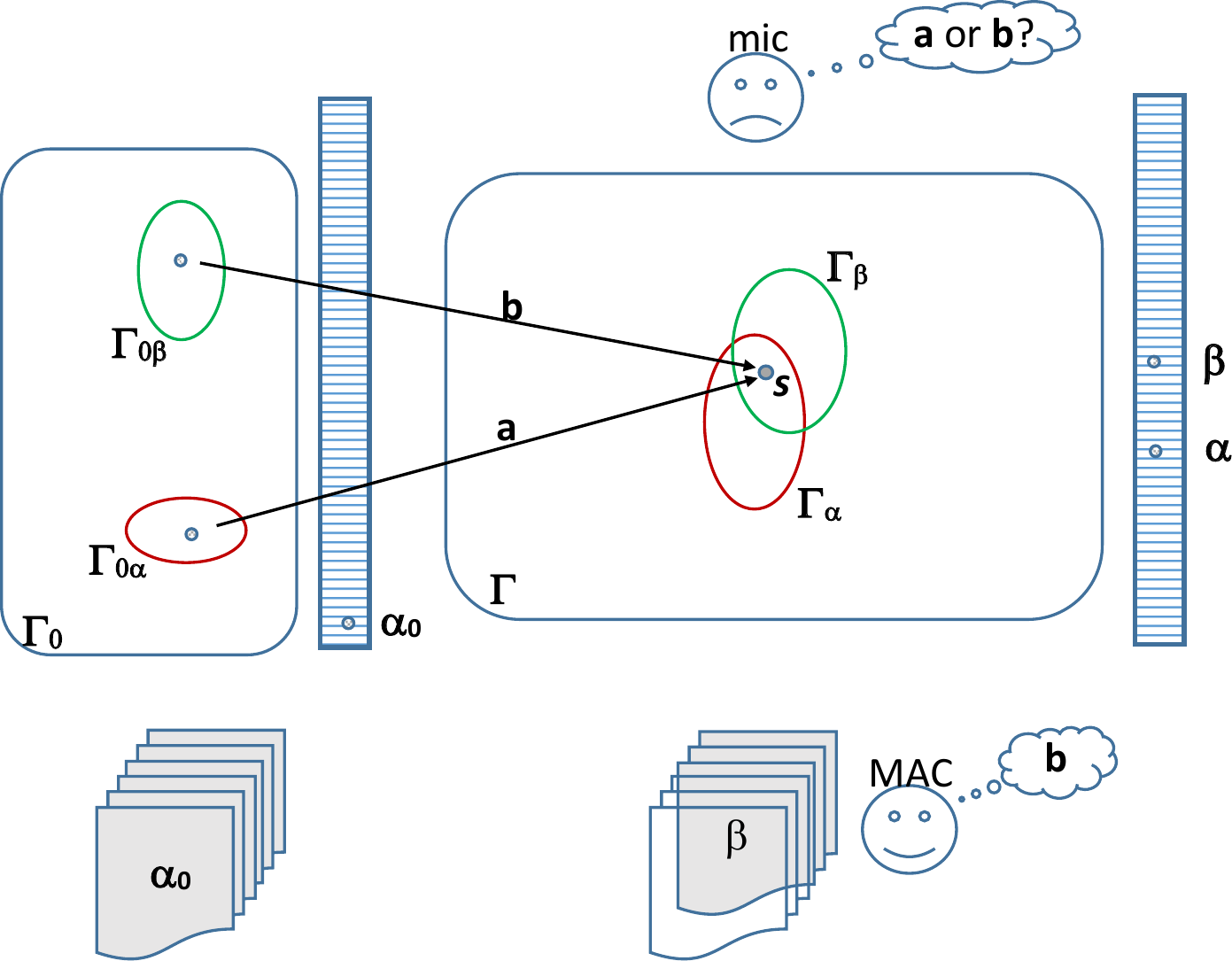}
\caption{Presentation of the system dynamics as a mapping between the initial, $\sti$, and final, $\stf$, state space of X. The initial and final state spaces of the system A+WR are presented too: its final state (e.g. $\alpha$, $\beta$) is the memory record. A "microscopic observer" knowing details of the final state space of X knows the final state $\st$ but cannot decide which history is true -- {\bf a} or {\bf b}. A "macroscopic observer" does not know  the final state of X but knows the value of the memory record, say, $\beta$.   This implies that "it" knows that the history is {\bf b} and the final (initial) state of X belongs in $\stfb$ ($\stib$).}
\label{fig:2}
\end{figure}

The relations (\ref{detevo}) imply that the initial state space of  X, $\sti$, may be split into disjoint regions $\stia$ so that if  $\st_0\in\stia$, the dynamics of the supersystem leads to the final memory record $\yw$. The set of all final values of states of X if the system begins at $\stia$ will be denoted as $\stfa$, which is a subset of the final state space of X, $\stf$. It is worth noticing that the final value  of X in $\stfa$ may not mean that the final memory record is $\yw$: specifically, the initial value of X that does not belong in $\stia$ may lead the system into a final state from $\stfa$ (the regions $\stfa$ may overlap, see Fig. \ref{fig:2}).

Now we introduce a special \emph{averaging} of the energy distribution over the state space of X at the beginning and the end of the process, $$F_0\equiv -\beta^{-1}\ln Z_0, \,\,\, F (\yw )\equiv -\beta^{-1}\ln Z(\yw ),$$ respectively, where $Z_0=\sum_{\sti} \e^{-\beta E_0(\st_0,\yw_0 )}$,  $Z(\yw )=\sum_{\stf} \e^{-\beta E(\st ,\yw )}$ and $\beta$ is an arbitrarily chosen positive factor. (We use the sum over state spaces, but a continuous description using the integrals may be used instead.)

We call it the \emph{free energy} in analogy with this concept in equilibrium statistical physics. Notice that the final free energy, $F=F (\yw )$, is defined for a concrete final memory record $\yw$. 

 Similarly, we define  free energies on regions $\stia$ and $\stfa$,
 $$\Fia \equiv -\beta^{-1} \ln Z_{0\yw} ,\,\,\,
 \Ffa \equiv -\beta^{-1}\ln Z_{\yw} ,$$
respectively, where $Z_{0\yw} = \sum_{\stia} \e^{-\beta  E_0(\st_0,\yw_0 )}$ and $Z_{\yw} =
 \sum_{\stfa} \e^{-\beta  E (\st ,\yw )}$.

The important fact concerning the regions $\stia$, $\stfa$ is that the mapping $\sti\to\stf$ realizing the dynamics of the supersystem is one-to-one on $\stia$.
The reason is that the mapping $(\st_0,\yw_0)\to (\st ,\yw )$ can be inverted since the dynamics is reversible. The second fact concerning the mapping $\stia\to\stfa$ is that each initial condition $\st_0$ leads to the same value of work $W_\yw$. 
We use these facts in deriving the fundamental identity, 
\begin{equation}\label{fi}
W_\yw = -\Delta \Ffa ,
\end{equation}
where $\Delta\Ffa\equiv \Ffa - \Fia$. Namely 
\begin{multline*}
\e^{\beta W_\yw}\e^{-\beta \Fia}=\\
=\sum_{\st_0\in\stia}\e^{\beta (E_0 (\st_0,\yw_0 ) - E(\st (\st_0,\yw_0), \yw (\st_0,\yw_0)))} \e^{-\beta E_0 (\st_0,\yw_0 )}=\\
=\e^{-\beta \Ffa }
\end{multline*}
(in a continuous case we use Liuville's theorem) and we get the identity
\begin{equation}\label{efi}
 \e^{\beta (W_\yw + \Delta \Ffa )} =1, 
\end{equation}
 which implies (\ref{fi}). It is valid for an arbitrary system X+A+WR whose states evolve in a deterministic way and the initial and final energy of the system X+A (including a possible interaction energy with WR) may be defined. The results then may be interpreted not only in classical but also in quantum mechanical description.  
 
 Though the free energy is used in the identity, it may not have anything in common with a thermodynamic description. 
  Nevertheless the equality (\ref{fi}) has an interesting thermodynamic meaning: a "macroscopic observer" who detects a concrete memory record obtains essentially more  precise information about the initial and final state spaces of the observed process, see Fig.\ref{fig:2}. The identity     (\ref{fi}) relates a difference of free energies on \emph{these} state spaces with the  work performed.

\section{Heat and thermodynamic equalities}
\label{sec:3}

While the work reservoir and auxiliary system play an exclusive role in our description, X is just a "rest of the universe" or some "surrounding matter", so that X+A+WR may be considered to be an isolated system. Nevertheless, our interest is to have X as a studied thermodynamic system.   
A thermodynamic system, however,  interacts not only with a work reservoir and possibly  with a control auxiliary device but  with  various heat reservoirs (heat baths). This may be  an essential supply of the system's energy (recall Planck's formulation of the Second Law) called the heat. 

If  a heat resevoir were included in X (see \cite{Tasaki2013}), i.e., the studied thermodynamic system were a subsystem Y of X, we would face a serious problem of differentiating the interaction of  Y  with this reservoir and the interaction of Y with A+WR at a microscopic level. The reservoir consists of microscopic subsystems that interact with the system during the time interval $(0,\tau )$, which leaves some memory records (realize that each change of a microscopic coordinate may be a memory record comparable with memory records of other microscopic systems, such as those of A+WR).

Some standard assumptions about a heat reservoir -- e.g. that it keeps no memory of the system's action \cite{Sekimoto2010} or fulfills related assumptions like self-equilibrating \cite{DefJar2013} -- may become problematic when studying the process at meso- or microscopic space and time scales. 
 
That is why we identify X with a studied thermodynamic system and define the heat exchange not as a new interaction within the time interval $(0, \tau )$  but as a random change of the system's initial state. It is  inspired by the work by G.E. Crooks \cite{Crooks1998}, where  the heat interaction is imagined as a random "jump" within the state space. Specifically, in Crooks' scheme the time evolution is related to a change of a control parameter $\lambda$. The heat exchange is represented by a constant $\lambda$ that may be interpreted as  an "instantaneous" change of the system state.  

In our approach, we thus imagine the process as consisting of two stages. During the first stage (before the time $t=0$), there is a pure heat exchange with a thermal environment that  adjusts the initial condition of X for the next stage. Since the interaction is random, the initial state of X, $\st_0$, is manifested with  a probability $p(\st_0)$.    

Then starts the second stage from  $t=0$ till $t=\tau$, during which the system interacts only with WR and A in the fully deterministic way described in the previous Section (this corresponds to the evolution of $\lambda$ in Crooks' scheme). Since system X is isolated from the heat reservoir during this stage, the process from $t=0$ till $t=\tau$ may be called the \emph{adiabatic process}.

The situation may be actualized experimentally when guaranteeing good isolation of the supersystem X+A+WR from the thermal environment during the second stage (and achieving an adiabatic process). In practice, however, this is a rather problematic task at a microscopic level because of the ubiquitous interaction of a microsystem with molecules in its surroundings (see the concept of microadiabaticity \cite{Marti2015}). Nevertheless, a suitable definition of  system X might be helpful here.

 The crucial difference between A+WR and X  consists in the possibility of adjusting the initial state of A+WR to a given value by an external manipulation at the beginning of the process.  Concerning the final state of A+WR, however, it is random because of the stochastic initial condition of X. Nevertheless, we can imagine a sophisticated dynamic interconnection between   A and WR so that  A returns to its initial state at the end of the process (the reset of memory kept in A). It is, however, not a trivial task to guarantee that this is possible. We call the set of final states of X for which it is possible \emph{adiabatically accessible} states in the analogy of this concept in macroscopic thermodynamics  \cite{LieYng1999}.

\begin{figure}
\includegraphics[width=0.4\textwidth]{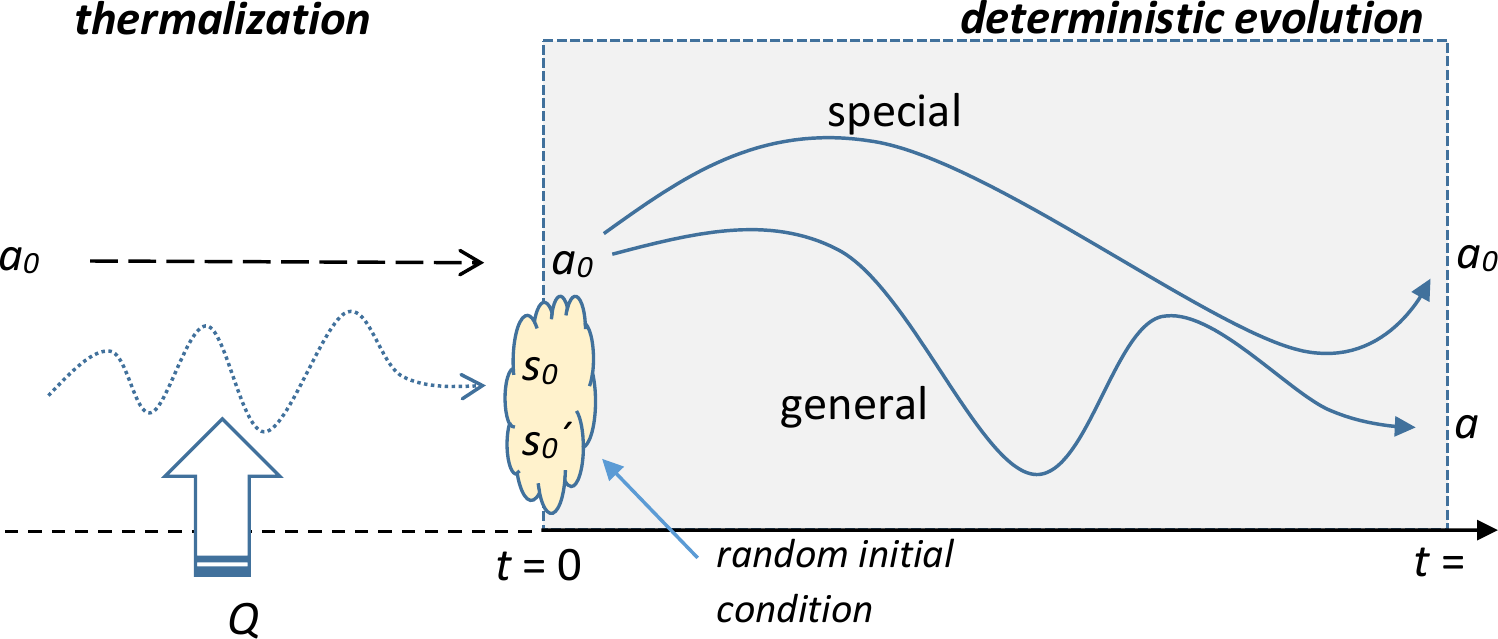}
\caption{Two stages of the process. In the first stage, the contact of the system X with a thermal environment adjusts random initial conditions at $t=0$. Then the system evolves in a deterministic way till $t=\tau$. The system dynamic is interconnected with the auxiliary device A and the work reservoir WR. If this interconnection is organized in such a special way that A returns to its beginning configuration, the process defines the adiabatic accessibility at microscopic level. }
\label{fig:3}
\end{figure}

The first stage has no time determination. It may be understood, for instance, as a sufficiently long interaction with a large heat bath in thermal equilibrium    
with the temperature $T=(\beta k_B)^{-1}$, where $k_B$ is Boltzmann's constant. It implies the canonical distribution of probabilities,
\begin{equation}\label{peq0}
p(\st_0)\equiv p^{eq}(\st_0) = Z_0^{-1}  \e^{-\beta E_0(\st_0,\yw_0 )} .
\end{equation}
There may also be various forms of thermal interaction during the first stage leading to a \emph{nonequilibrium}
  probability distribution at the time $t=0$. 

If we imagine the first stage of the process (thermalization) as a random "jump" and  $\tau$ as a sufficiently small time interval, we can realize the thermodynamic process in  Crooks' spirit as many repetitions of these two stages. During this process, the work is defined  as a result of the  interaction with a work  reservoir and not via an external control parameter $\lambda$.   
  
Since the process is deterministic after the time $t=0$ (during the second stage), the probability of the final state at $t=\tau$ with  the  memory record  $\yw$ is 
\begin{equation}\label{pa}
 p_\yw = \sum_{\st_0\in\stia} p(\st_0 ). 
\end{equation}
The final value of the memory record is then a random function that is correlated with the initial condition of  system X. The mutual information  between X and A+WR at the end of the process is in this deterministic (error-free) arrangement \cite{SagUed2013}: 
 \begin{equation}\label{mi}
I=-\sum_\yw p_{\yw}\ln p_{\yw}=- \av{\ln p},
\end{equation}
where the averaging of  a quantity $x (\yw )$ is defined as 
$$\av{x}\equiv \sum_\yw p_\yw x(\yw ).$$ 
The identity (\ref{fi}) then may be presented in the averaged form. However, to get a more common form of this identity, we introduce the difference of free energies 
$\Delta \GF\equiv \Ffa - F_0$. Then we can write (\ref{fi}) in the form 
\begin{equation}\label{gsu}
W_\yw = -\Delta\GF +J_\yw ,
\end{equation}
where $J_\yw\equiv \Fia - F_0 \ge 0$. 
Averaging gives 
\begin{equation}\label{agsu}
\avdE = -\av{\Delta \GFav}+\av{J} .
\end{equation}
If the initial state of X is in thermal equilibrium 
with the temperature $T=(\beta k_B)^{-1}$, the probability of finding the initial state in $\stia$ is 
\begin{equation}\label{peq}
p_{\yw} \equiv p_{0\yw}^{eq} = Z_0^{-1} \sum_{\stia} \e^{-\beta E_0(\st_0,\yw_0 )} .
\end{equation}
Then 
$$\av{J}=-k_B T\sum_\yw p_{0\yw}^{eq}\ln p_{0\yw}^{eq} = k_B T I.$$
Hence, (\ref{agsu}) becomes the generalized Second Law 
\cite{SagUed2008,SagUed2009,SagUed2010,Sag2011}
and $\av{\Delta \GFav }$ is then the change of nonequilibrium free energy \cite{ParHorSag,EspBro2011}.

The equality (\ref{agsu}) thus plays the role of the generalized Second Law even in the case of a nonequilibrium initial state. 
Since $\av{J}\ge 0$, we get the inequality   
\begin{equation}\label{gine}
\avE \le \av{\Delta\GFav},
\end{equation}
where $\avE\equiv -\avdE$ is the averaged change of the energy of the system X+A.  The quantity $\av{\Delta\GFav}$ thus plays the role of a \emph{maximal average energy}  that a work reservoir can \emph{supply} into its surrounding. This result  is valid for an arbitrary initial thermodynamic state of a rather general physical system X. Nevertheless, while its interpretation in the framework of classical physics is straightforward, its meaning for a quantum meachanical system is not studied here (see the recent results concerning quantum processes \cite{FaiRen2018}).

\section{Second Law at a Longer Time Scale}  
 \label{sec:4}

The generalized Second Law (\ref{agsu}) says that the average energy (work) supplied \emph{from}  a work reservoir into a system being in thermal equilibrium at the beginning of the process is limited by the difference of nonequilibrium free energy $\av{\Delta \GFav}$. In classical thermodynamics, on the contrary, there is an upper limit to the average energy that can be transformed \emph{into} a work reservoir from a system starting at an equilibrium state: it cannot be \emph{larger} than $-\Delta F$, where $\Delta F\equiv F - F_0$ is a standard thermodynamic difference of free energies.    

To compare previous results with the consequences of classical thermodynamics, we rewrite the identity (\ref{fi}) in the form  
\begin{equation}\label{hol}
W_\yw = -\Delta F (\yw ) +D_\yw ,
\end{equation}
where $\Delta F(\yw )\equiv F(\yw )- F_0$ and  $$D_\yw\equiv -\beta^{-1}\ln \frac{p_{0\yw}^{eq}}{p_{\yw}^{eq}} $$ 
 with 
 $$p_{\yw}^{eq}\equiv Z(\yw )^{-1} \sum_{\stfa} \e^{-\beta E(\st, \yw )}.$$

Let the initial states of X occur with probabilities $p_{0\yw}^{eq}$ corresponding to the equilibrium state set by the contact with a heat reservoir having the temperature $T$. Then the  average of the term $D_\yw$, 
$$\avD\equiv -\sum_\yw p_{0\yw}^{eq}\ln \frac{p_{0\yw}^{eq}}{p_\yw^{eq}}\le \ln (\sum_\yw p_\yw^{eq}),$$
and we see that if $\stf_\yw$ are disjoint sets, then $\sum_\yw p_\yw^{eq}\le Z(\yw )^{-1} \sum_{\stf} \e^{-\beta E(\st, \yw )}= 1$, and $\avD \le 0$. 

Imagine now the situation in which the final state of the auxiliary device is the same as its initial  state, i.e.  $a=a_0$, independently of the initial state of the system X, i.e., we suppose that all final states of X are adiabatically accessible (see previous Section). That is, the process from $t=0$ to $t=\tau$ includes also the return of the auxiliary device to its initial state (thanks to a sophisticated dynamic interconnection between A and WR and a suitable choice of the time $t=\tau$; nevertheless it is not clear if it is always possible to actualize such an arrangement).   The essential memory record is thus  only the state of the work reservoir alone. 

Now $E(\st ,\yw )=E(\st ,a_0,\wre )\approx E(\st, a_0,\wre_0)=E(\st ,\yw_0)$ if the interaction energy between WR and A+X at the final state is negligible. Then the final free energy does not depend on $\yw$ and $F(\yw )=F(\yw_0)\equiv F$. If $\stf_\yw$ are disjoint sets, then $\avD \le 0$,  which implies  the validity of the classical thermodynamic equality $\avdE \le - \Delta F$. 

If, however, there are processes  starting at states from different regions $\stia$ and finishing at the same state $\st$ (see Fig. \ref{fig:2}) then $\avD$ may be positive. This is possible only if the final memory records of these processes are different (the memory records -- states of the work reservoir -- play the crucial role). 

If $\avD >0$, the situation seems to contradict the Second Law though it is in  agreement with the information thermodynamics: there is an information coupling (correlation) between the system X and the system WR. This correlation, however, cannot be reset: the state of WR belongs to the thermodynamic cycle. To clarify the situation, we will study a thermodynamic process in which the two stages (thermalization and deterministic evolution) are repeated many times.   

First, the system X is in thermal contact with a thermostat of the temperature $T$ and its state space is $\sti$. Then a process with an arbitrary auxiliary system A (e.g. including measurement and feedback) is realized. Suppose that $\Delta F=0$, A returns to its initial state  and positive work $W_\yw$ is performed (supplied into the work reservoir). The probability of doing that is $p_{0\yw}^{eq}$. 

At this moment, the system X is in a nonequilibrium state. Then it is put into contact with the same thermostat to reach  equilibrium with the same temperature as at the beginning (during this process it may spontaneously pump the heat energy from the thermostat). The thermodynamic cycle is thus finished. During this process, the state of the memory register (i.e. the work reservoir) is kept at $\yw$. Then it is isolated from the thermostat and a new cycle may begin. Thanks to the equilibration, $p_{\yw}^{eq}$ means now the probability that the system is at $\stfa$.

Let us denote $\stfa^{inv}$ as the region of state space in which the states are the time-reverse of states from $\stfa$: if $\st =(x,v)$, the operation of time reverse, $\trev (\st )$, gives $\trev (x,v) = (x,-v)$, where $v$ represents the velocity coordinates. If the energy of the state $\st$ is the same as that of $\trev (\st )$, the system after the equilibration can be found in $\stfa^{inv}$ with the probability $p_{\yw}^{eq}$. The identity (\ref{hol}) implies that   
\begin{equation}\label{trans}
 p_{\yw}^{eq} = p_{0\yw}^{eq} \e^{\beta W_\yw } .
\end{equation}
Assume now that the time reverse of the memory register, $\yw$, is the same value,
\begin{equation}\label{trev}
 \trev (\yw ) = \yw 
\end{equation}
 (e.g. if the weight in the gravity field is characterized only by its position and the initial state of A includes no velocities). This means that if the system is in $\stfa^{inv}$, its next evolution is a time reversion of a trajectory from $\stia$ to $\stfa$ and the energy $W_\yw$ \emph{returns}  from the work reservoir into the system X.  

The crucial result coming from (\ref{trans}) is that if $W_\yw >0$, the probability of a reverse process that returns the energy from the work reservoir is \emph{larger} than the probability that positive energy is transformed into the work reservoir. This has very important consequences. To explain them, let us imagine a long thermodynamic process in which the cycles defined above are permanently repeated.   
 
Due to the random exchange of energy during the first stages, the whole process is stochastic. In other words, there is an infinity realizations of it. Let us imagine one such a concrete realization $r$, in which the work reservoir changes its position at the end of each cycle to form a succession: $\wre_0, \wre_1, \ldots$. Let the probability of reaching the value $\wre_i$ during the $i$-th cycle be $p^{(i)}$. The probability of this concrete realization is thus
 \begin{equation}\label{prob}
 P(r)= p^{(1)}p^{(2)}\ldots .
 \end{equation}  
During this realization there are "antithermodynamic"  cycles in which the energy of WR increases, $\wre_{j+1} - \wre_j\equiv \Delta \wre_j > 0$. Denote the sum of all such $\Delta \wre_j$ as $W_+$. 
 
 Now imagine a different realization, $r_{rev}$,   
 that is longer than $r$ since it includes the reverse cycle after each "antithermodynamic" cycle. The  probabilities  of other ("thermodynamic") cycles in this realization remain the same. The probability of the realization $r$ is lower than that of $r_{rev}$ (though $r_{rev}$ is longer than $r$). Namely,
 \begin{equation}\label{probrev}
 P(r)= P(r_{rev})\e^{-\beta W_+}.
 \end{equation}  
The probability of a realization during which the accumulation of energy essentially exceeds $\beta^{-1}T$ is thus extremely low in comparison to the probability of another realization in which this accumulation is canceled. An average of the performed work over all realizations, $\overline{W}$, cannot be positive.

 The system behaves -- in a larger time scale --  in agreement with the fluctuation theorems \cite{Sev2008} and equation (\ref{trans}) expresses the transition rates between two states of the work reservoir with different energies. Regardless of a control of the system by any Maxwell's demon, the states of the work reservoir fluctuate and cannot accumulate an usable energy. This corresponds to the solution of the puzzle of Maxwell's  demon in the spirit of M. Smoluchowski \cite{Smol1912} and R. Feynman \cite{FLS1966}. 
 
 The result may seem to contradict the fact that the averaged work $\avdE$ is positive (since $\avdE =\avD >0$). But $\avdE$ arises as an average of many cycles starting at the \emph{same} initial state of WR, $\wre_0$, while $\overline{W}$ is an average over successions of cycles starting at \emph{varying} initial states. The fact that $\overline{W}$ and $\avdE$  may differ so dramatically indicate that there is a  dynamic link between the system and the work reservoir so that the behavior of the system is sensitive on the value of the initial state of WR (see next Section).     
  
We see that regardless of how sophisticated the control of the system used to transform a positive work into the reservoir in one thermodynamic cycle is, the performed work may be canceled in the next cycle so that the probability of an accumulation of positive work is negligible low. The result has been derived, however, under the assumption (\ref{trev}), meaning that the time reverse of the final state of the work reservoir as well as the final state of the auxiliary device is \emph{static}, i.e., their time reverses are the same states.

The situation when it is \emph{not true}  appears in \cite{ManJar2012}: A system having three states ($A,B,C$) is dynamically connected with a single bit having two states ($0,1$). The bit represents the  auxiliary device in our scheme. When a process $C\to A$ occurs, heat is
withdrawn from the thermal reservoir to lift the mass in the gravity field (work reservoir). This process, however, may happen only if the transition of the bit $0\to 1$ occurs simultaneously. At this moment, we have a positive energy gain in WR but also a change of the state of the auxiliary device since $a\neq a_0$. To reset the state of the bit into the initial state $a_0=0$, a transition $A\to C$ has to happen simultaneously (because of its dynamic interconnection) and the work gained is returned into the system and the thermal reservoir. When, however, there is not one bit but a sequence of bits in a \emph{movement} (a stream of bits), the  bit at the state $1$ may be replaced by the bit at the state $0$ as a result of this movement without a loss of energy in the work reservoir. 

This is the case of an autonomous system that may essentially accumulate heat energy in the work reservoir during a long time period. The movement of the auxiliary system (a sequence of bits) is crucial (see similar systems studied in \cite{ManQuaJar2013,LuManJar2014}).  The process may be understood as a  permanent writing of information into the bit sequence that thus plays the role of a memory register. This information-reservoir approach \cite{BoManCru2016,BarSei2014,BarSei2014b,
ShiMatSag2016} may have important applications in molecular biology  \cite{HorSagPar2013,BaSei2013}. 

Our analysis of the validity of the Second Law concerns situations when (\ref{trev}) holds. However, it is not a generally valid condition (e.g.  the velocity of the piston in a Szilard engine cannot be ignored, see \cite{HatSas1998}), and, moreover, in many hypothetical arrangements it is not clear if it is valid or not (see e.g. \cite{Abr2012}, where an important part of a specifically  modified Szilard engine  -- a movable cylinder -- is in fact a memory register that may  keep the information in a form of its velocity  at the end of the cycle). Concerning an autonomous half-working Szilard engine (see Fig. \ref{fig:1}), an auxiliary device has to be rather complicated  to guarantee an actualization of the whole cycle. That is why we present the analysis of a simple system resembling in a way the half-working Szilard engine without any specification of an  auxiliary device (the  condition (\ref{trev}) is not used).

\section{A simple barrier model}
 \label{sec:5}
 A half-working Szilard engine has two different operating cycles. The engine either performs a given piece of work (when the molecule is in the left part) or does not work at all (when it is in the right one), see Fig. \ref{fig:1}. The following simple example has the same property.  

\begin{figure}
\includegraphics[width=0.4\textwidth]{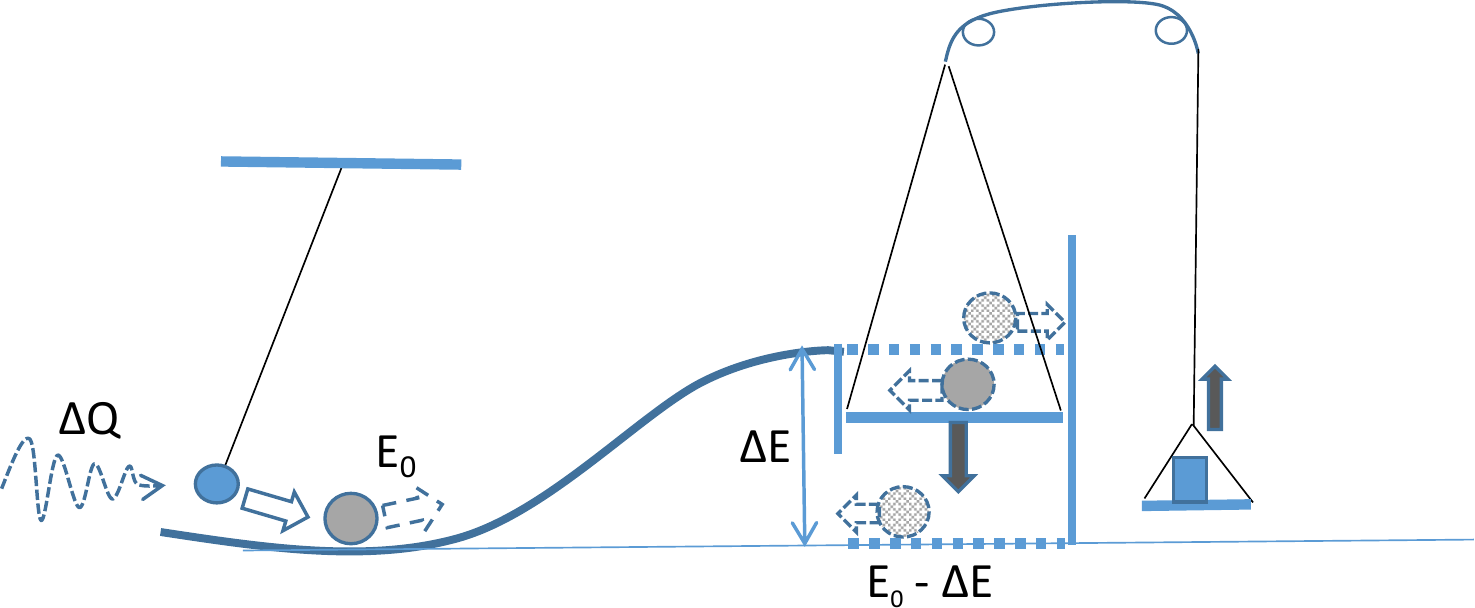}
\caption{A toy mechanical model of an extracting barrier. A thermalized pendulum is the source of energy; a ball (at $t=0$ at rest) obtains an energy $E_0$, goes up and transmits a part of its  energy to a lift raising a weight if  $E_0$ exceeds $\Delta E$. The transfer of energy to the ball is random, so the final position of the lift is random too: either it is at a height corresponding to the energy $\Delta E$ or is at the zero position (no shift).}
\label{fig:4}
\end{figure}

Consider  a system without degeneracy and with equidistant energy levels. The states may be labelled with natural numbers, i.e. $s_0=x_i$, and we may write $E_0(x_i)=i\epsilon$, $i=0,1,\ldots$.  At $t=0$, the contact with a heat bath is closed and the system is at a state $x_i$ with the probability $p(x_i)=Z_0^{-1}\e^{-\beta i\epsilon}$, where $Z_0=\sum_{i=0}^{\infty}\e^{\beta i\epsilon}=(\e^{\beta\epsilon} -1)^{-1}$. The process from $t=0$ to $t=\tau$ is realized as follows. If the energy of the state is less than a given energy threshold $\Delta E$, i.e. $E_0(x_i)<\Delta E =n\epsilon$, the system remains at the same state $x_i$. If the energy equals or is larger than the threshold, the energy $\Delta E$ is extracted from the system into a work reservoir and its energy changes to $E_0(x_i)-\Delta E$. It resembles an arrangement in which a particle may cross over a potential barrier via thermal activation \cite{FleHan1993,AsfShi1993}. When crossing this barrier, the potential energy of the  particle reaches the value $\Delta E$. Suppose that it may be used to raise a mass in the gravity field (see Fig. \ref{fig:4}).

 The probability that the system transfers the energy $\Delta E$ into the work reservoir is $$p=Z_0^{-1}\sum_{i=n}^{\infty}\e^{-\beta i\epsilon}=\e^{-\beta\Delta E}.$$
 This implies that the averaged work,
\begin{equation}\label{avEspec}
\av{W} =\e^{-\beta\Delta E}\Delta E,
\end{equation}
is positive and reaches its maximal value $\e^{-1}k_BT$  if $\Delta E=k_BT$.   

To understand this transfer of energy, it is important to say \emph{what the final state} of the system  is when the energy $\Delta E$ is extracted. We concentrate on two possibilities:\\
a) there are new states $x_j '$ in which the system appears after the extraction,\\
b) final states remain in the initial state space.\\[1mm]
\noindent
a) The state $x_i$ becomes the state $x_j '$ with the energy $E(j)=E_0(x_i) -\Delta E$. We thus get again the equidistant energy levels  so that $j=i-n$. Denoting $$G=\sum_{i=0}^{n}\e^{\beta i\epsilon}$$ 
we have $Z=Z_0+G$. The difference of free energies at the final and initial states is 
\begin{equation}\label{diffF} 
\Delta F=-k_BT\ln (1+\frac{G}{Z_0})<0.
\end{equation}
On the other hand, $p_{w_0}^{eq}+p_{w}^{eq}=1$ what implies that $\avD\le 0$. This means that the positive gain of energy is only due to the expansion of the state space (the decrease of the free energy). The gain of energy has thus the same essence as that during the expansion of a gas. The thermodynamic cycle, however, is not finished -- to do it we must perform a compression of the state space which needs some supply of energy.\\[1mm]
\noindent
b) In this case, the transfer of energy is organized in a way in which the system returns to a state $x_k$ even if it supplies a part of its energy into a work reservoir (if it crosses the barrier).  That is, the state $x_i$ either remains the same (if $i<n$) or it is transferred into the state $x_{i-n}$ with lower energy. Thus  
a complete thermodynamic cycle arises in which $\Delta F=0$. The averaged extracted energy is again (\ref{avEspec}), i.e., it is positive during a complete thermodynamic cycle.

The case b) seemingly contradicts the Second Law. The mutual information between the system and the work reservoir, $I=-p\ln p -(1-p)\ln (1-p)$, is nonzero, which implies that there is a correlation between these two systems. This causes the effect of a net gain of work. Specifically, the position of the weight is the memory record in the system's surroundings keeping  information about the extraction process. 

Let us study the process b) over a longer time period, i.e., when performing more than one cycle. Considering  the situation when the final state after the first cycle, $x_k$, is under the energetic barrier, i.e. $k<n$, there are two possibilities of reaching this final state:  $$(x_k,w_0)\to (x_k,w_0),\,\,\,\,{\rm or}\,\, \,\,\, (x_m, w_0)\to  (x_k, w ),$$
where $m=k+n$. Notice that the position of the weight keeps the memory of which possibility has occurred. The first possibility means that there was
 no crossing of the barrier and the system remains in its initial state. 
   
In the next step, we assume the same dynamics, i.e.,  the state $x_k$ (under the barrier) remains the same during the next step regardless in which state the work reservoir is, i.e. that the dynamics must include the possibility $$(x_k, w)\to (x_k\,w).$$ 
However, this means that the state $(x_k, w)$ of the supersystem X+WR is a final state of two different initial states, $(x_k, w)$ and $(x_m,w_0)$. This is a contradiction with a deterministic time reverse. 
 
It implies that the second cycle cannot be realized in the same manner as the first cycle. In other words, the states under barrier are not \emph{stable} in the longer time period because the transition $(x_k,w )\to(x_k,w)$ is excluded for any $k<n$. This means that a dynamic link between the system and the work reservoir is so strong that the change in the work reservoir ($w_0\to w$) essentially changes the dynamics of the system X. 

To be able to guarantee the stability of the low-energy states (and thus keep the main features of the system  dynamics demanded), we must consider a more complex arrangement in which there is some new degree of freedom, say $a$, of an auxiliary external system A.  It is necessary to assume a change of $a$ during the process in which there is no change in the work reservoir, i.e.
$$(x_k,w ,a_0)\to (x_k,w ,a ).$$ 
This implies that it is necessary to establish an auxiliary process to keep the stability of states  under the barrier. 

When introducing a new external parameter $a$, the premise of the Planck formulation of the Second Law is not fulfilled: specifically, there is some more external change (a record)  in a system's surroundings. As a result, there is no violation of the Second Law.  
Our simple model thus reveals a possible deeper connection of the Second Law with some fundamental restrictions on  dynamics that are not straightforwardly visible from the main results of information thermodynamics.

\section{Conclusions}
 \label{sec:6}
We study a thermodynamic process in this arrangement: during the first stage the system X is thermalized via a random exchange of energy with a heat reservoir. The next stage is a deterministic evolution during which the system exchanges energy with the work reservoir WR and interacts also with an auxiliary system A (see Fig. \ref{fig:3}). An autonomous supersystem X+A+WR  is an isolated dynamic system during the second stage. The final state of  WR+A is a memory record  $\yw$ of the concrete evolution of X. 
 
A concrete memory record $\yw$  may be understood as information for an external "macroscopic observer" for whom the detailed evolution of X is not visible. This information restricts the possible set of initial, ${\stia}$, and final, ${\stfa}$, states of X of such a process (see Fig. \ref{fig:2}). The difference of free energies defined on these sets determines the  work performed (\ref{fi}). This identity includes the generalized Second Law in a form of equality and other equalities, even in a more general form concerning the situation when both the final and initial states are not in thermal equilibrium.  
 
   We concentrate on processes in which the work reservoir is the only memory record of the process  and the initial state of the second stage is the thermodynamic equilibrium. Here the solution of the paradox of Maxwell's demon by information thermodynamics is insufficient:  the memory cannot be reset since the change in the work reservoir belongs to the thermodynamic cycle. If there is a positive gain of the average work, $\avdE >0$, the situation seems to be a violation of the Second Law.
 
 Nevertheless, the analysis of this situation over a longer time period, i.e. when the system repeats many cycles, shows that an accumulation of the positive work gain is impossible (or extremely improbable) regardless of how sophisticated the control of a Maxwell demon managing a positive gain of work is (see \cite{EaNo1999}, pp. 19-20). 
  Namely, during the interaction with a heat reservoir (when the system may spontaneously draw energy from its surroundings) the state of the system randomly "wanders" over its state space. The system may finish this wandering in a state that is a time reverse of
a final state of a process $\stia\to\stfa$.

  If the condition (\ref{trev}) holds, the system then performs a time reverse, i.e. a process $\stfa\to\stia$. If the work gain in WR, $W_\yw$, is positive, the reverse process means a loss $-W_\yw$ in WR.  Formula (\ref{trans}) implies that if $W_\yw$ is positive, the probability of a reverse process is higher. In a longer time period including many cycles, this means that an accumulation of positive work in WR is practically impossible, since the "antithermodynamic" cycles are canceled at longer time scales  with a high probability and the system behaves in accord with the fluctuation theorem (compare \cite{EvaDeb1994,Wang2002}). 
 
In other words, the system that fulfills the condition $\avdE >0$ (a positive supply of energy into the work reservoir in average) may not be able to accumulate a positive energy in work reservoir during many repeating of the cycle, i.e. $\overline{W}\le 0$. The reason is a dynamic interconnection between the work reservoir and the system: the dynamic evolution of the system is sensitive on the initial state of the work reservoir. It has a direct relation to the condition (\ref{trev}) since it indicates a certain instability of the whole autonomous system when WR has a higher energy (the weight is shifted up).

 If the condition (\ref{trev}) does not hold, the situation is not so clear.  A nice example of this situation is a stream of bits introduced in \cite{ManJar2012}.  It shows that such a system may accumulate positive work during a longer time period. Here, however, the device processes the information reservoir (incoming stream of bits) into positive work in accord with information thermodynamics (an information flow exists \cite{HorEsp2014}) .  
  Since a single bit plays the role  of an auxiliary device (that may control the system) in our arrangement, some  relations of the  "measuring-feedback"  and  "information-reservoir" approaches  \cite{BarSei2014} may be studied further.     
(Notice also that the relation of the time $\tau$ -- duration of the cycle -- and the typical time when a bit changes into the next one in the sequence then gives various modes of operation of the system \cite{ManJar2012,ManQuaJar2013}.) 
 
 The analysis of the last example studied here does not use time-reverse states or the condition (\ref{trev}). The simple system representing an energetic barrier shows that there may be pure dynamic reasons that prevent to actualizing a "demonic" system. The sensitivity of the system on the initial state of the work reservoir is revealed again. Here, however, without using an idea of a backward running of the system. Some new external structures interacting with the system and changing their state during the process must be introduced to guarantee the  behavior demanded (stability of low-energy states). The final state of these structures then plays  the role of a memory record and the system works in accord with the results of information thermodynamics.  This single result obtained on an extremely simple model opens some additional questions concerning a possible deeper connection of dynamics and thermodynamics.  
 
Notice also that the very existence of cycles in which all memory records but that in WR  are reset is deeply connected with the condition of adiabatic accessibility \cite{LieYng1999}. Since this is  more dynamic than thermodynamic condition (contrary to the concept of (micro)adiabatic process \cite{Marti2015}) it opens again questions concerning the role of dynamics  in thermodynamics of meso- or microscopic systems.

\section*{Acknowledgements}

The author is very indebted to Jan Minar for critical reading of the manuscript  and stimulating discussions. The work is supported by the New Technologies Research Center of the West Bohemia University in Pilsen.

\end{document}